\documentclass[a4paper,10pt]{elsarticle}

\usepackage[utf8]{inputenc}
\usepackage{graphicx}
\usepackage{float}
\usepackage{hyperref} 
\usepackage{isotope} 
\usepackage{upgreek}
\usepackage{sidecap}
\begin{document}

\begin{frontmatter}

\title{Pulse-shape discrimination of surface events in CdZnTe detectors for the COBRA experiment}

\author[Dresden]{M.~Fritts\corref{cor1}} \ead{matthew\_christopher.fritts@tu-dresden.de}
\author[Dortmund]{J.~Tebr\"ugge} 
\author[Muenchen]{J.~Durst} 
\author[Hamburg]{J.~Ebert} 
\author[Dortmund]{C.~G\"o\ss{}ling} 
\author[Dresden]{T.~G\"opfert} 
\author[Dresden]{D.~Gehre} 
\author[Hamburg]{C.~Hagner} 
\author[Hamburg]{N.~Heidrich} 
\author[Dortmund]{M.~Homann} 
\author[Dortmund]{T.~K\"ottig} 
\author[Dortmund]{T.~Neddermann} 
\author[Hamburg]{C.~Oldorf} 
\author[Dortmund]{T.~Quante} 
\author[Dortmund]{S.~Rajek} 
\author[Dresden]{O.~Reinecke} 
\author[Planck]{O.~Schulz} 
\author[Hamburg]{J.~Timm} 
\author[Hamburg]{B.~Wonsak} 
\author[Dresden]{K.~Zuber} 

\address[Dresden]{Technische Universit\"at Dresden, Institut f\"ur Kern- und Teilchenphysik \newline Zellescher Weg 19, 01068 Dresden}
\address[Dortmund]{Technische Universit\"at Dortmund, Lehrstuhl f\"ur Experimentelle Physik IV \newline Otto-Hahn-Str~ 4, 44221 Dortmund}
\address[Hamburg]{Universit\"at Hamburg, Institut f\"ur Experimentalphysik \newline Luruper Chaussee 149, 22761~Hamburg}
\address[Muenchen]{Ludwig-Maximilians-Universit\"at M\"unchen, Faculty of Physics \newline Schellingstr~4, 80799~M\"unchen}
\address[Planck]{Max-Planck-Institut f\"ur Physik, Werner-Heisenberg-Institut \newline Foehringer Ring 6, 80805~M\"unchen}
\cortext[cor1]{Corresponding author. Tel 49 351 46334568; fax 49 351 46337292.}

\date{}

\pdfinfo{%
  /Title    (COBRA lateral surface event discrimination)
  /Author   ()
  /Creator  ()
  /Producer ()
  /Subject  ()
  /Keywords ()
}


\begin{abstract}
Events near the cathode and anode surfaces of a coplanar grid CdZnTe detector are identifiable by 
means of the interaction depth information encoded in the signal amplitudes. However, the amplitudes
cannot be used to identify events near the lateral surfaces. 
In this paper a method is described to identify lateral surface events by means of their pulse shapes.
Such identification
allows for discrimination of surface alpha particle interactions from more penetrating forms of radiation,
which is particularly important for rare event searches.
The effectiveness of the presented technique in suppressing backgrounds due to alpha contamination in the search for neutrinoless double beta decay with the COBRA experiment is demonstrated.
\end{abstract}



\end{frontmatter}

\section{Introduction: The COBRA experiment}
The aim of the \underline{C}admium Zinc Telluride \underline{0}-Neutrino Double \underline{B}eta Decay \underline{R}esearch \underline{A}pparatus  (COBRA) experiment \cite{cobra} 
is to prove the existence of neutrinoless double beta decay by investigating the decay of the isotope \isotope[116]Cd, with a sum energy of both electrons of 2.814 MeV. 
The experiment employs CdZnTe semiconductor detectors of the coplanar grid design. A DAQ system records complete signal forms, so that pulse shape \hyphenation{ana-ly-sis}analysis is possible. 
Details of the system can be found in \cite{schulz}.
An array of detectors is currently operating at the Gran Sasso Underground Laboratory (LNGS) in Italy.
 A complete description of the experiment can be found in \cite{cobra_ahep_2013}.

The target decay has a very long half-life of more than $10^{25}$\,yr\cite{rodejohann}. For this reason the expected count rates are very low. 
Consequently, a critical issue for COBRA is to reduce background, both by minimizing the detection of unwanted background signals and by identifying those that are detected via some signature.  
Identified background events can thus be vetoed in the analysis.

In the current low-background setup, the main component of background signals in the energy region of interest stems from alpha radiation 
from \hyphenation{con-ta-mi-na-tion}contamination on the detector surfaces and surrounding hardware such as holder structures and shielding layers. As the range of alpha particles is very small, 
those events are limited to the surface of the detectors, in contrast to double beta decays, which occur homogeneously distributed throughout the detector volume.

\section{Coplanar grid principles}

\begin{SCfigure}
 \centering
 \includegraphics[width=0.5\textwidth]{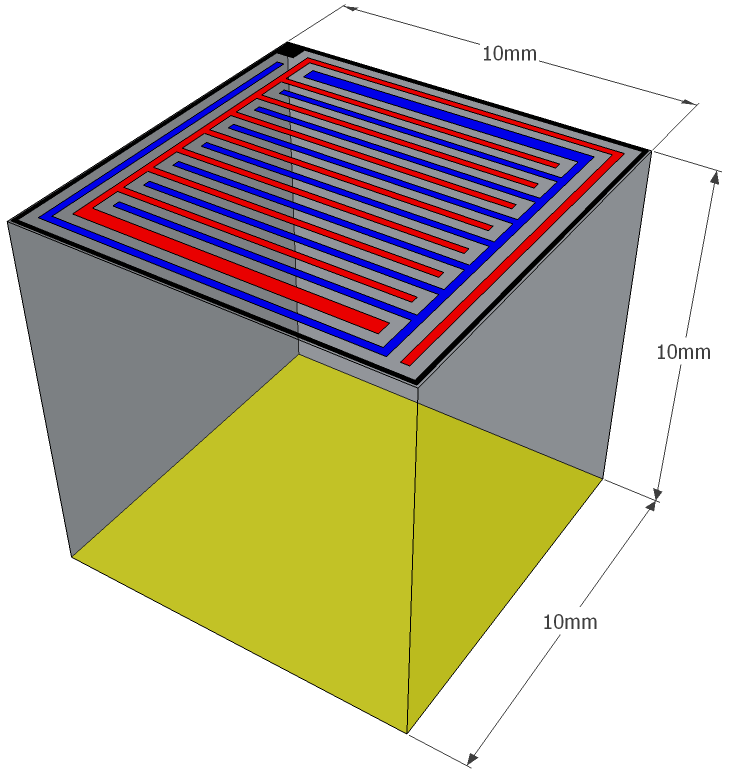}
 \caption{Schematic of CPG detector of the type used in the COBRA experiment. The coplanar anode grids are indicated in red and blue. In addition a metalized guard ring surrounds the grid, shown in black. The cathode is shown in yellow on the opposite side.}
 \label{fig:cpg}
\end{SCfigure}

Due to the poor transport properties of holes in CdZnTe, the coplanar grid (CPG) technology was proposed by Luke\cite{cpg94}, in analogy to the Frisch grid in gas ionization chambers. On one face of a cuboid crystal, 
the anode is segmented into two comb-shaped interlocking grids, which are set to different electric potentials (see figure~\ref{fig:cpg}). This potential difference is referred to as the grid bias. 
One anode, called the collecting anode (CA), is held at ground potential; 
the other one, called the non-collecting anode (NCA), is at a negative potential of typically $-60$\,V. 
On the opposite face the cathode is held at a negative bias near $-1.2$\,kV. 
Signals read out from each anode record the induced charge from drifting charge carriers freed by a particle interaction at a sampling rate of 100\,MHz. 
The values given here correspond to the current operation of the COBRA experiment. 

\begin{figure}
 \centering
 \includegraphics[width=1.00\textwidth]{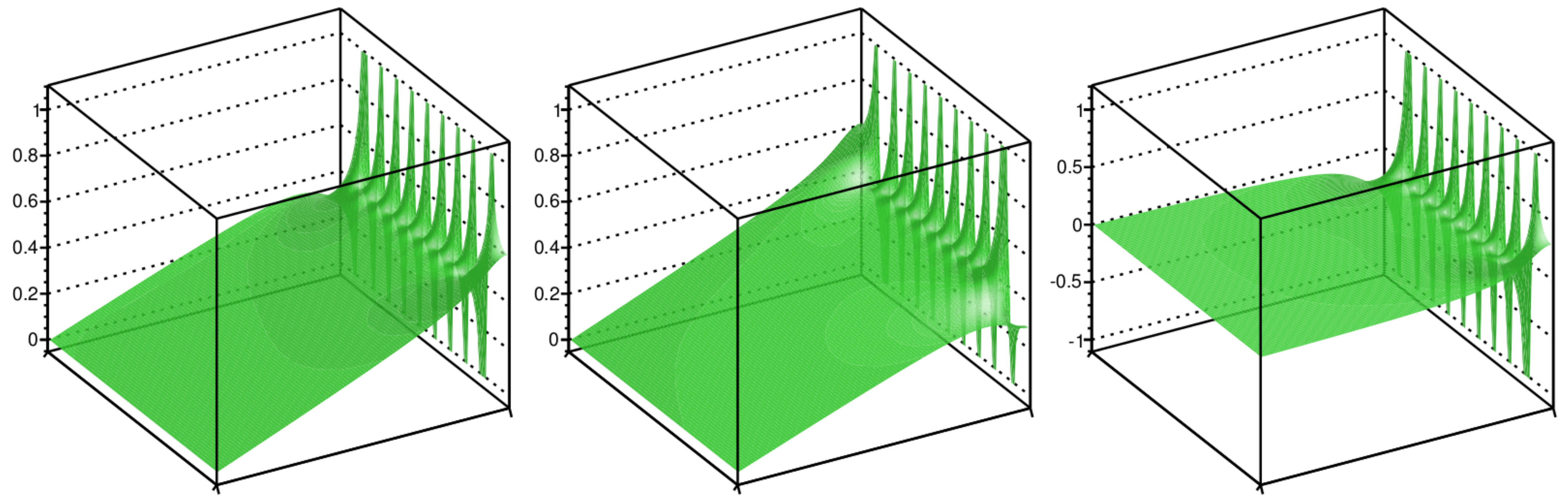}
 \caption{Weighting potentials along a cross section of a CPG detector, 
calculated from the electrode geometry. The anode grids are on the right side in each figure. Left: CA weighting potential. Middle: NCA weighting potential. Right: Difference weighting potential.}
 \label{fig:wpots}
\end{figure}

In accordance with the Shockley--Ramo theorem the dimensionless quantity known as the weighting potential can be calculated for the electrode configuration.
The weighting potentials of both anode grids rise nearly uniformly toward the anode throughout most of the detector volume. 
Close to the anodes, they diverge into a peak-and-valley structure, with the peaks terminating on the strips of the corresponding anode. The peaks in the CA weighting potential are thus located at the same position as the valleys in the NCA weighting potential, and vice versa. It is instructive to also define a difference weighting potential, $U_{\textrm{diff}}$, constructed by subtracting the weighting potential of the NCA from that of the CA. $U_{\textrm{diff}}$ is approximately 
zero throughout the bulk of the detector, diverging near the anode plane to maxima at the CA strips and minima at the NCA strips. The calculated weighting potentials are illustrated in figure~\ref{fig:wpots}.

\begin{SCfigure}
 \centering
 \includegraphics[width=0.7\textwidth]{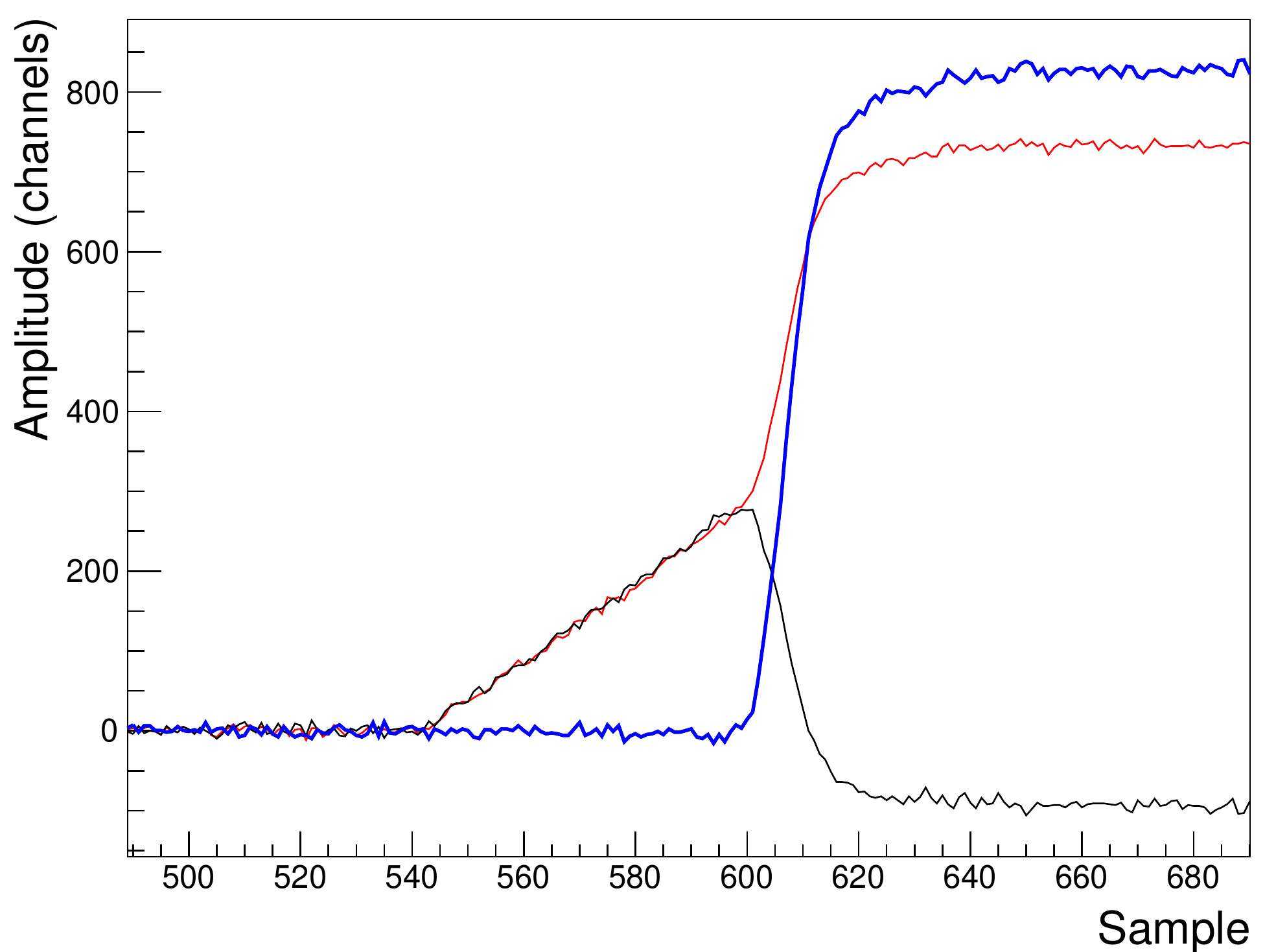}
 \caption{Pulses from a typical event. Red: CA pulse. Black: NCA pulse. Blue: Difference pulse. 
The period between samples 540 and 600 corresponds to the charge cloud drifting through the detector bulk, after which
it enters the region near the anode plane.
The event shown is a 2.0\,MeV gamma interaction from a \isotope[228]Th calibration of a detector operating in the LNGS setup. The
sampling rate is 100 MHz.
}
 \label{fig:centpulse}
\end{SCfigure}

The corresponding pulse shapes produced by a drifting electron cloud can be understood by considering the path of the cloud through the weighting potentials. The signals on the NCA and CA rise equally as the electron cloud drifts through the 
bulk of the detector toward the anodes. When entering the area of the grid bias influence, the charge cloud drifts toward a collecting anode strip. This results in a sharp rise and fall of the CA and NCA pulses, respectively. 
The difference pulse, defined as the NCA pulse subtracted from the CA pulse, will be nearly zero until the electron cloud enters the near-anode region, at which point it will rise sharply to a maximum, forming an approximate step function shape.  
Figure \ref{fig:centpulse} shows pulses from a typical event.

The energy deposition of an incident particle can be calculated from the difference between the CA and NCA pulse
 amplitudes. The resolution is greatly improved by applying a weighting factor to this difference to compensate for significant 
electron trapping (equation~\ref{eq-energy}). The weighting factor $w$ is determined empirically 
for each detector. The amplitudes $A$ and weighting factor can also be used to 
calculate the interaction depth of the ionizing particle, defined as the 
distance from the anode plane normalized to the cathode-anode 
distance (equation~\ref{eq-depth}). Details concerning energy and depth reconstruction 
based on the pulse amplitudes can be found in \cite{FrittsCPG}. The depth information along the longitudinal axis of the 
detector allows for the identification of surface events on the cathode and anode surfaces, but not the other four surfaces 
(the lateral surfaces).
\begin{equation}
E \propto A_{CA} - w \cdot A_{NCA} 
\label{eq-energy}
\end{equation}
\begin{equation}
z = \frac{1+w}{1-w} \cdot \ln \left( 1 + \frac{1-w}{1+w} \cdot \frac{A_{CA} + A_{NCA}}{A_{CA} - A_{NCA}}  \right)
\label{eq-depth}
\end{equation}

In terms of the basic CPG principle an ideal anode grid would have an infinitesimal pitch, so that the final rise/fall of the signals would be extremely sharp (apart from effects due to the finite size of the charge cloud). In practice the pitch is limited to a minimum dictated by leakage current effects. Thus some additional structure appears in the final part of the pulse shapes. Moreover the structure will depend on the specific position of the interaction relative to the anode grid pattern. 
If the charge cloud drifts in a straight line toward a CA strip, the difference pulse will rise relatively smoothly, as the charge path follows a peak structure in the difference weighting potential. In contrast, a charge cloud which starts its drift along a path pointing to an NCA strip will be diverted toward a CA strip near the anode plane. In the process it can be expected to pass through a valley structure in $U_{\textrm{diff}}$ before entering a peak structure as it approaches a CA strip. 
The resulting difference pulse will thus fall somewhat below the baseline before sharply rising to the final maximum value. Such variations in pulse shape are a normal feature of events occurring in the bulk of the detector.

\begin{SCfigure}
 \centering
 \includegraphics[width=0.6\textwidth]{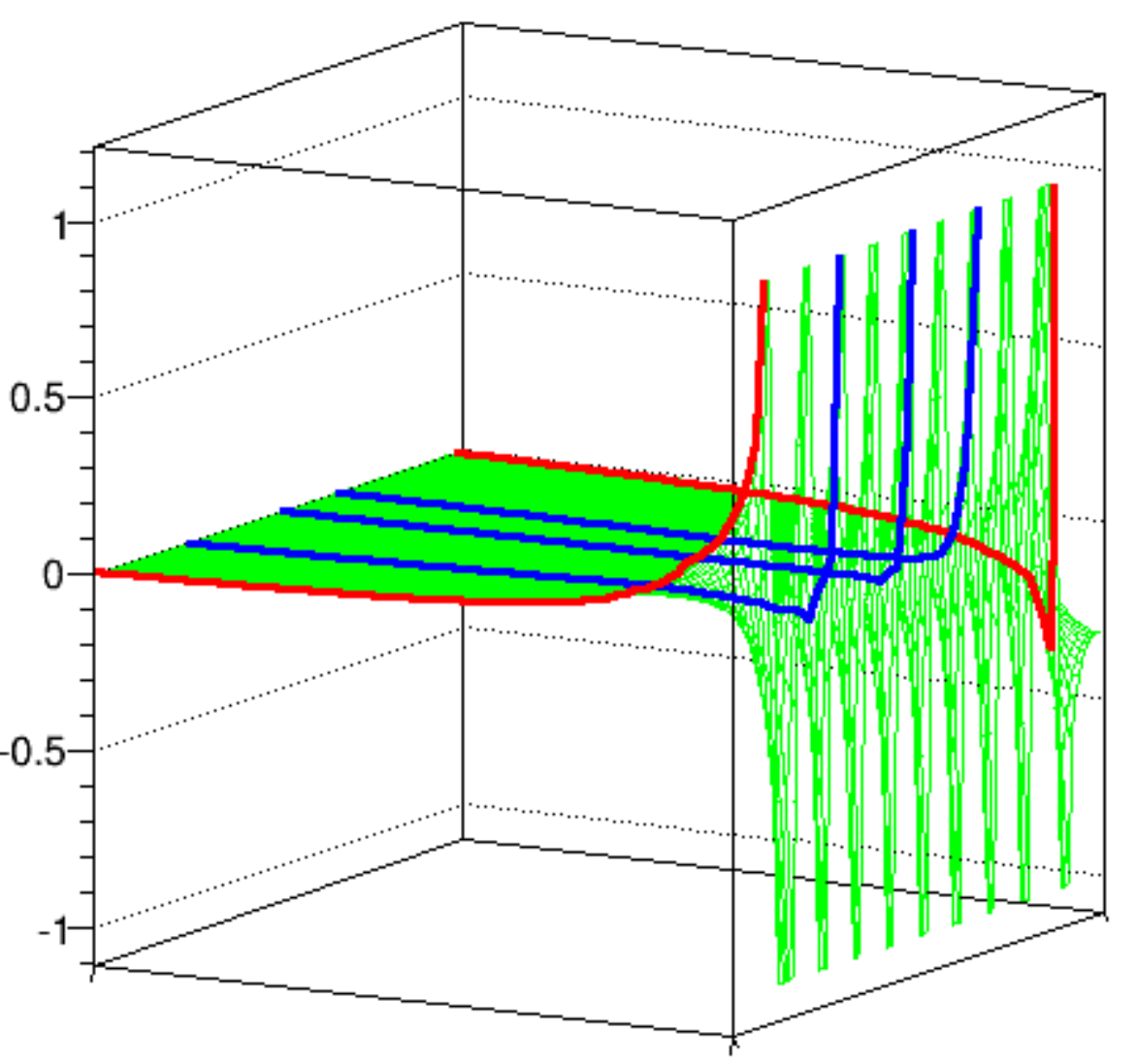}
 \caption{Difference weighting potentials along charge paths from interactions in the bulk (blue) and near the lateral surfaces (red).}
 \label{fig:diffwp-paths}
\end{SCfigure}

Near the lateral surfaces, stronger distortions in the weighting potentials occur. Thus for interactions occurring near a lateral surface, the variations in pulse shape are stronger. If the interaction occurs near a lateral surface adjacent to an outer CA strip, the difference pulse will rise more smoothly 
than would be seen for an interaction occurring in the bulk. For interactions near a lateral surface adjacent to an outer NCA strip, the difference pulse will fall unusually far before rising to its final value. These distortions thus provide a means for identifying lateral surface events (abbreviated as LSEs). Figure~\ref{fig:diffwp-paths} compares the difference-pulse shapes expected from events in the bulk and near the lateral surfaces by plotting the $U_{\textrm{diff}}$ values corresponding
to the charge paths. 

It should be pointed out that the extent of the lateral distortions will depend on the specific grid design and operation. The COBRA CPG detectors currently employed in the 
low background setup are manufactured by EI Detection \& Imaging Systems (formerly eV Products)~\cite{endicott} in a design which includes an additional guard ring electrode along the edges of the anode plane, 
surrounding the anode grid. The guard ring was originally devised to reduce lateral distortions in the weighting potential~\cite{grid-improvement}. However for simplicity of operation the guard rings are left floating in the COBRA setup, which eliminates the smoothing effect.  For CPG detectors with a guard ring held at a fixed potential, the distortions in pulse shapes from lateral surface events are expected to be smaller. 
The distortions should also be smaller in grid patterns with no guard ring, in which the grids extend closer to the edges of the anode plane.

\section{Experimental setup to explore surface events}

To investigate the distortions in pulses from lateral surface events (LSEs), a laboratory setup has been built to irradiate a detector with an alpha-emitting radioactive source. The source contains \isotope[241]Am with an activity of 330\,MBq. The prominent alpha energies are 5486\,keV and 5443\,keV with relative intensities of 84.5\% and 13.0\%, respectively. Since the particles must travel through several millimeters of air, the detected energies are significantly smaller. 
The very small penetration depth of alpha particles in CdZnTe (less than 20\,$\upmu$m) means that they will always produce surface events.
Two gamma radiation sources are also used to provide a sample of events uniformly distributed in the detector. 
A \isotope[137]Cs source produces gammas at
662\,keV which are used for the energy calibration of the detector, and a \isotope[232]Th source emits gammas 
at several energies, up to 2615\,keV, for better comparison with high-energy alpha events. The three sources can 
be used together in the same run or separately.

\begin{figure}
\center
 \includegraphics[width=0.9\textwidth]{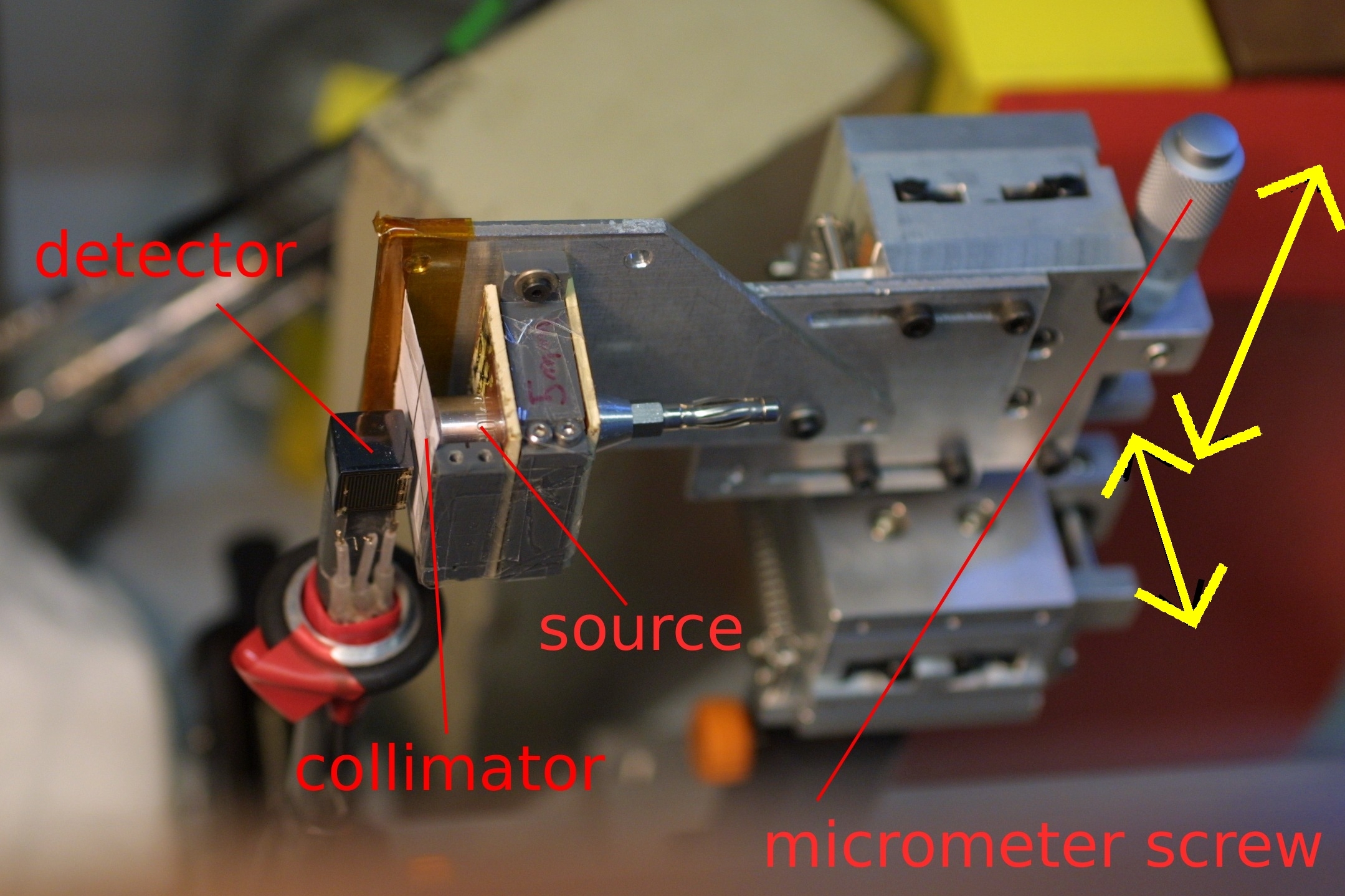}
 \caption{Experimental setup for irradiating the detector with alpha particles. The yellow arrows indicate the directions in which the alpha source and collimator can be moved to target a specific spot on the detector surface. The gamma sources are not visible.}
 \label{pic:experimental-setup} 
\end{figure}
\begin{figure}
 \includegraphics[width=0.9\textwidth]{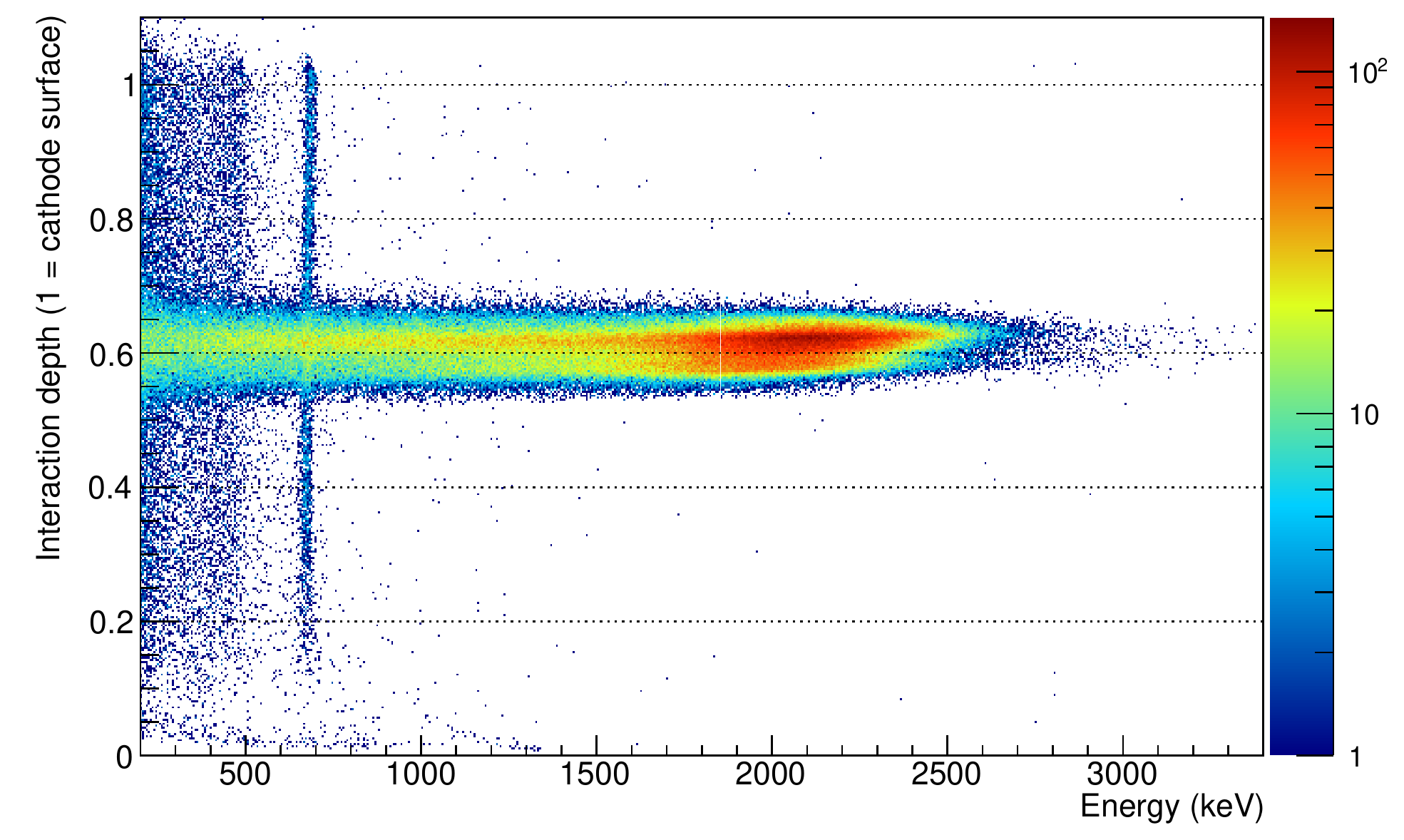}
  \caption{Calculated interaction depth along the longitudinal (anode-to-cathode) axis plotted against the deposited energy for a data run from the laboratory setup. Visible are the Compton continuum and the vertical line of full energy deposition of 662\,keV gamma radiation from the \isotope[137]Cs source, homogeneously distributed within the detector, and the alpha peak up to 2.5\,MeV concentrated at an interaction depth of 0.6.}
 \label{fig:interact-depth-beam-diameter}
\end{figure}

The experimental setup is shown in figure~\ref{pic:experimental-setup}. A collimator is used to provide a tight beam of alpha particles so 
that a specific spot on the surface of the detector can be targeted. The collimator is a sheet of paper with a small hole in it fixed at 
0.5\,cm from the source. The mounting holding the source and collimator can be moved along two perpendicular axes with micrometer screws in approximately 0.01\,mm steps. Figure~\ref{fig:interact-depth-beam-diameter} shows the calculated interaction depth
along the longitudinal axis versus energy
for a run in which the alpha and \isotope[137]Cs gamma sources are used. By definition the depth is normalized to the anode-to-cathode distance of the detector, which is 1\,cm. Thus
from such plots we can estimate the diameter of the alpha beam on the 
detector surface to be approximately 1\,mm. The detector is mounted so that the
alpha beam targets a lateral surface. Whether this surface is a CA or NCA side can be chosen without moving the detector by choosing which of the two anodes to hold at ground potential. 

%

\section{Quantifying pulse shape distortions}

\begin{figure}[h]
 \centering
 \includegraphics[width=1.0\textwidth]{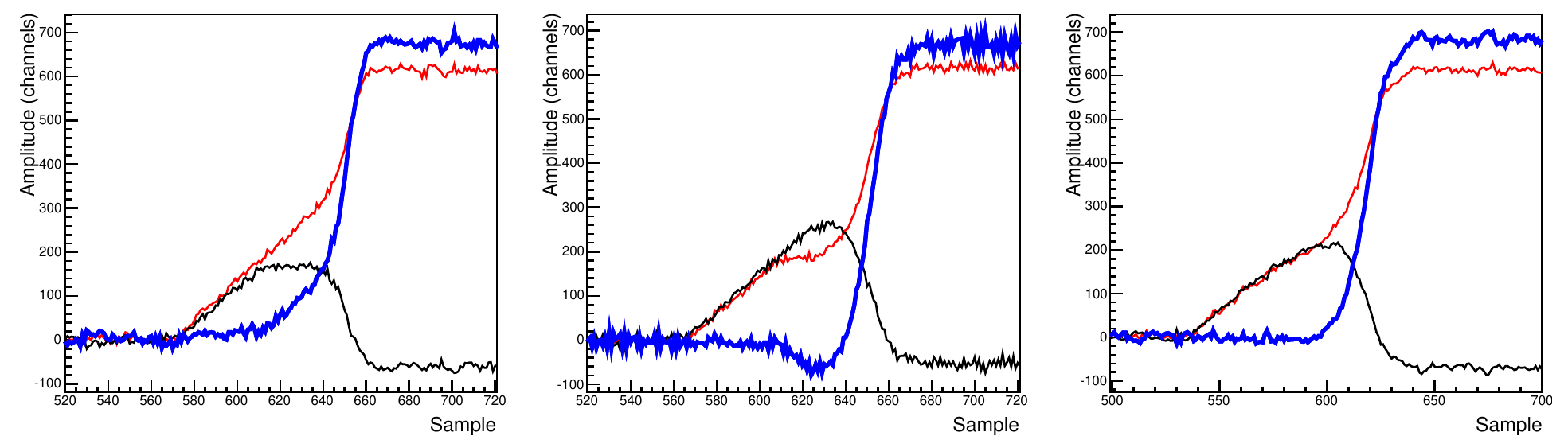}
 \caption{Pulses from a CA-side surface event (left), an NCA-side surface event (center), and a central event (right). 
Each event has an energy of 1.9\,MeV.
The color code is the same as in figure~\ref{fig:centpulse}.}
 \label{fig:distpulses}
\end{figure}

Figure~\ref{fig:distpulses} shows typical pulses resulting from alpha particle interactions at a CA surface and at an NCA surface. For comparison typical pulses from a gamma particle interaction are also shown. 
The three events are chosen to have very similar energies and interaction depths along the longitudinal axis, so that differences in the pulse shapes are primarily due to their positions in the plane parallel to the anode grids. 
The surface event pulses show characteristic distortions compared with the central event pulses,
similar to those depicted in figure~\ref{fig:diffwp-paths}.

To develop a discrimination tool for LSEs based on pulse shape analysis, it is useful to define quantities to characterize the pulse distortions. Because the characteristic distortions for CA- and NCA-side LSEs differ qualitatively, 
two different quantities have been defined. In both cases the analysis is performed on the difference pulse, which shows the distortions most dramatically. 

For an LSE on the CA side, the characteristic distortion in the difference pulse is a slow rise relatively early in the pulse, followed by the standard fast rise. 
We define a quantity referred to as the early rise time (ERT). This is the time in samples (1 sample = 10 ns) between the points at which the pulse rises from 3\% to 50\% of its final height. 
These points are chosen to target only the first half of the rising portion of the pulse, since the 
timing of the second half depends on other effects such as diffusion of the charge cloud. 
The 3\% point is found by searching backwards from the 50\% point to minimize the influence of pre-pulse fluctuations.
 ERT is characteristically larger for CA-side LSEs than for central events. 
Due to the quick rise of the difference pulse the resolution of ERT is limited by the sampling rate. 

\begin{figure}[h]
 \centering
 \includegraphics[width=1.0\textwidth]{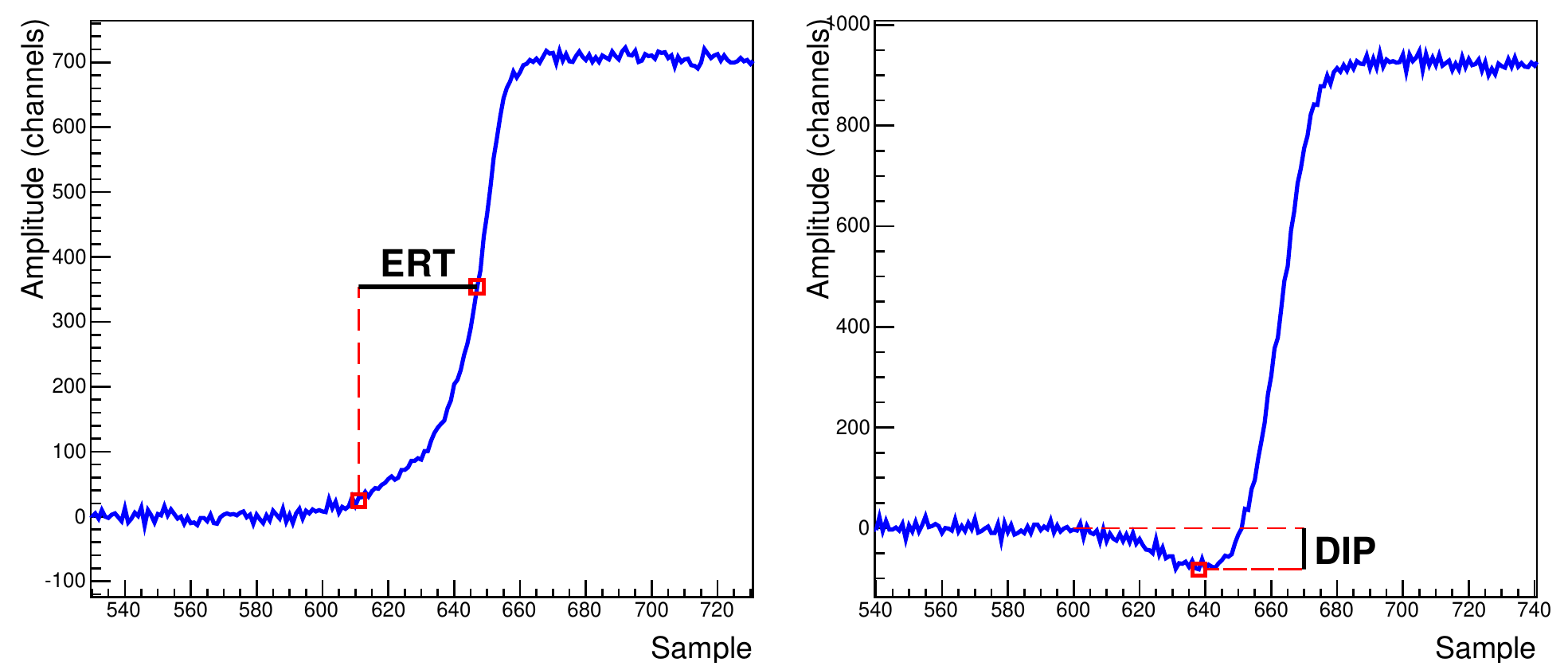}
 \caption{Illustration of definitions of LSE identification quantities ERT (left) and DIP (right) from the 
shape of the difference pulse.}
 \label{fig:defs-ertdip}
\end{figure}

For an LSE on the NCA side, the difference pulse falls significantly below the baseline before the final rise of the pulse. 
We define a quantity known as the dip of the pulse (DIP). 
This is the maximum amount by which the value of the pulse falls below the baseline.
To minimize the influence of pre-pulse fluctuations, the range for DIP is restricted to a window 30 samples wide with the right edge at the 50\% 
point of the pulse. Due to this limited window the value of DIP can be negative in pulses where the relevant distortion is not present.
DIP is characteristically larger for NCA-side LSEs than for central events. 
The definitions of the quantities ERT and DIP are illustrated in figure~\ref{fig:defs-ertdip}.

Both of the quantities describing LSE distortions have been defined so that for each 
a simple threshold can be set to discriminate LSEs from central events. For simplicity of analysis it is desirable that a fixed threshold can be set 
that applies for all energies. At low energies, where pulse shape analysis is more likely to fail due to poor signal-to-noise ratios, it is desired that an LSE cut be conservative, i.e.\ that by default events are categorized as central events. Hence both ERT 
and DIP have been defined such that they tend to be small for low-energy events. For high-energy LSEs, the two quantities behave differently with energy: ERT is roughly constant with energy, while DIP is roughly proportional to energy. 
This is because DIP is essentially an amplitude quantity, whereas ERT is a timing quantity primarily independent of the number of charge carriers.
A different definition of DIP so that it is normalized to the event energy is possible, but it would not have the desired conservative behavior for 
low-energy events.

\begin{figure}[h]
 \includegraphics[width=\textwidth]{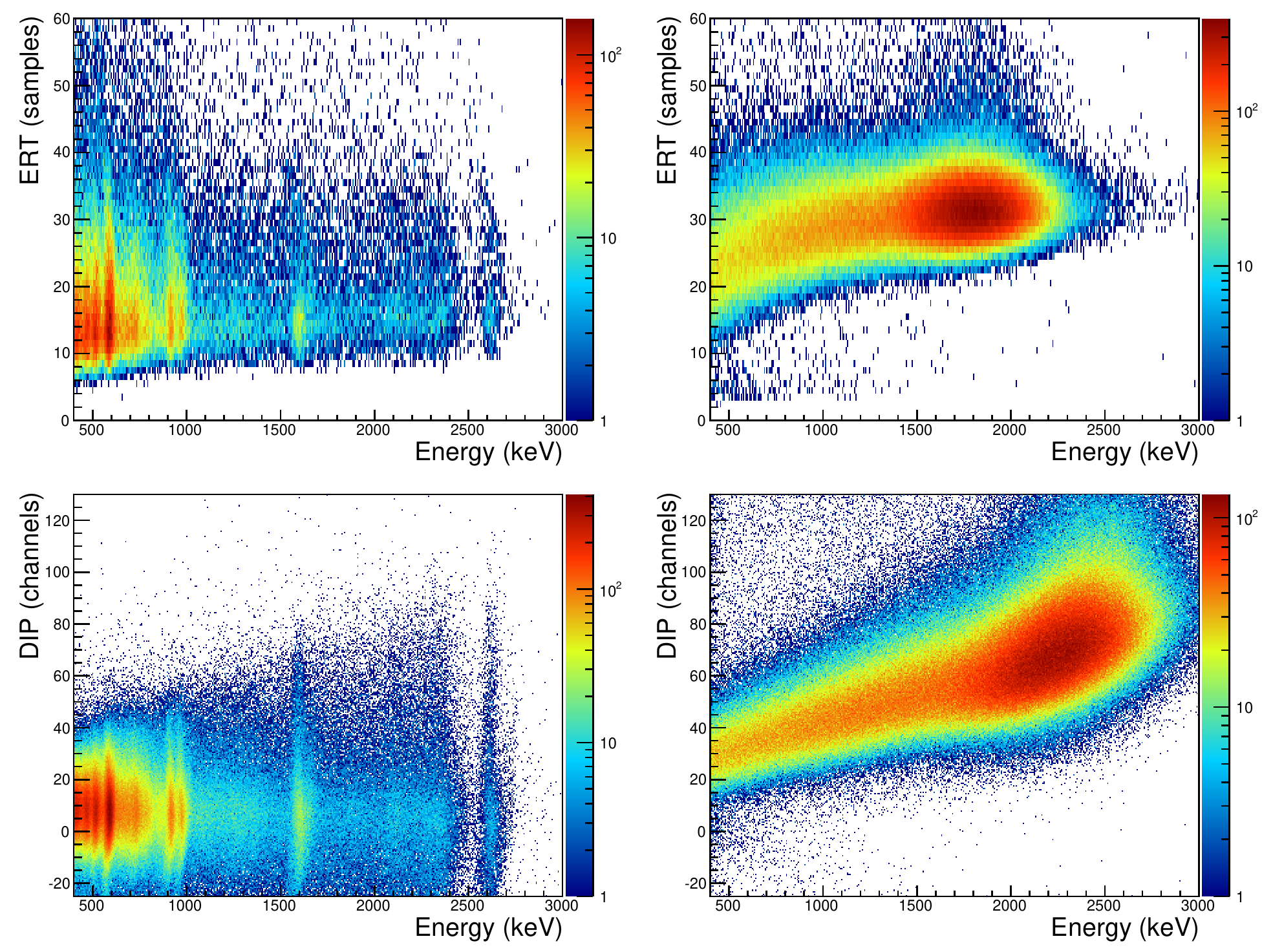}
 \caption{LSE identification quantity versus energy for various radiation source configurations. 
Upper Left: ERT, \isotope[232]Th gamma source. 
Upper Right: ERT, alpha beam directed near the center of a CA side. 
Lower Left: DIP, \isotope[232]Th gamma source. 
Lower Right: DIP, alpha beam directed near the center of an NCA side. } 
 \label{fig:ertdipVSe}
\end{figure}

Figure \ref{fig:ertdipVSe} shows how the two quantities behave for gamma particles and for alpha particles targeting
the corresponding side of the detector. 
Events near the anode surface have been removed from these plots to avoid near-anode distortions typical to CPG detectors~\cite{FrittsCPG}.
The gamma population, provided by the \isotope[232]Th source, consists of full-energy peaks up to 2615\,keV along with a continuum of Compton-scatter events.
Most of these uniformly-distributed events fall in characteristic bands for both LSE identification quantities with very little energy dependence over the full energy range. 
 It is clear that both quantities take on a characteristically higher value for the alpha population than for the gamma population, 
and that a cut that rejected events above a certain threshold value would remove most of the alpha population (all LSEs) while retaining most of the gamma population (mostly central events).
 

\begin{figure}[h]
 \includegraphics[width=1.0\textwidth]{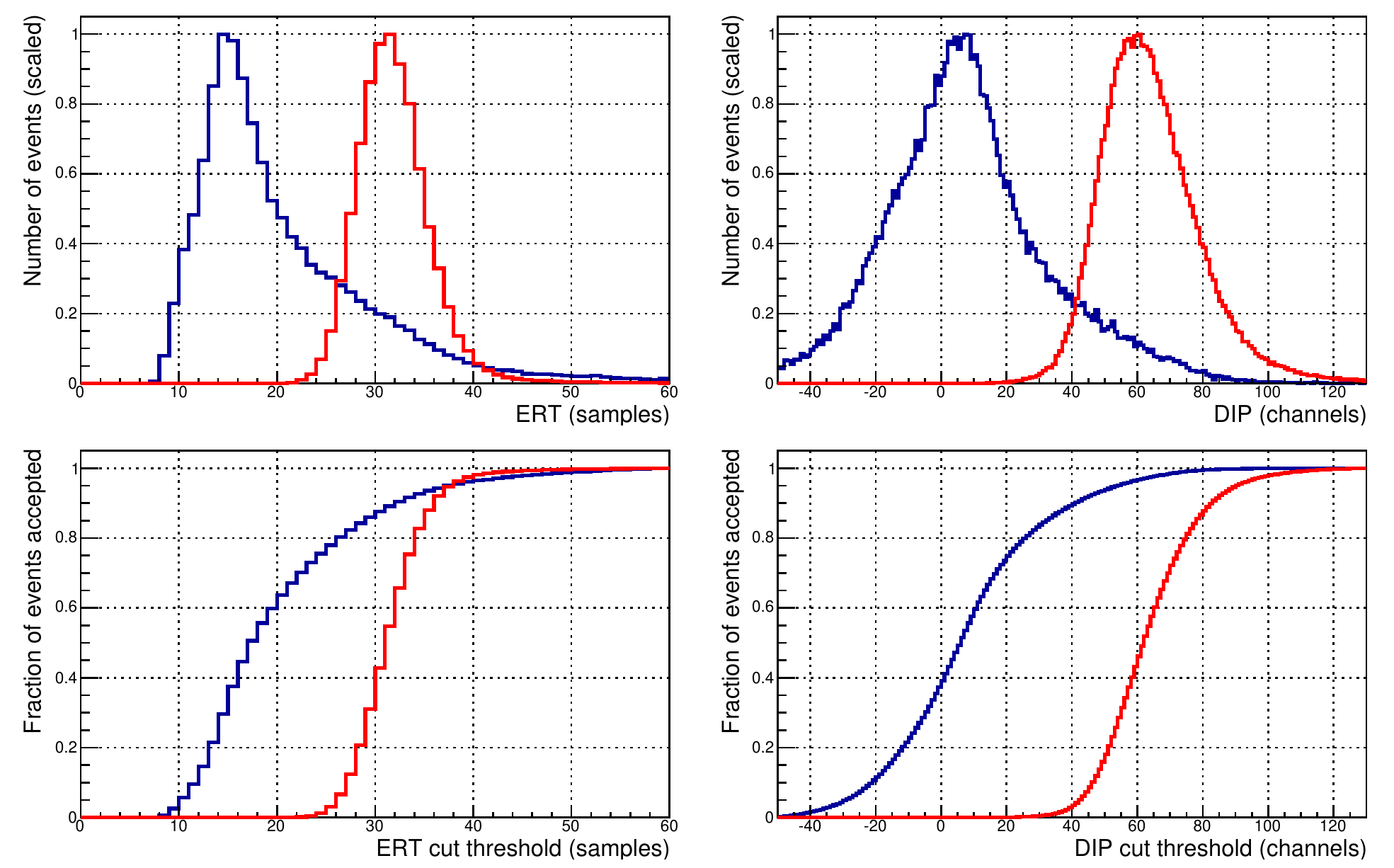}
 \caption{Distribution of LSE identification quantities for gammas (blue) and alphas (red) in an energy range
of 1.7-2.0\,MeV. 
Upper Left: ERT distributions (alpha events on CA-side). 
Upper Right: DIP distributions (alpha events on NCA-side). 
Lower Left: Fraction of events accepted as a function of ERT cut threshold.
Lower Right: Fraction of events accepted as a function of DIP cut threshold. } 
 \label{fig:cutDiscrim}
\end{figure}

In figure~\ref{fig:cutDiscrim} the distributions of the quantities for gammas and alphas are compared for a specific energy
range to better illustrate the discrimination power. For example, a cut with an ERT threshold of 26 would remove more 
than 90\% of the CA-side alphas while sacrificing 21\% of the gammas. A cut with a DIP threshold of 46 would remove 
more than 90\% of the NCA-side alphas while sacrificing 8\% of the gammas. Combining both cuts would sacrifice
29\% of the gammas. Note that these percentages are specific to this situation, and may vary with 
the location on the detector surface of the alpha interactions, as well as with detector and energy range.  

\begin{SCfigure}
 \centering
 \includegraphics[width=0.7\textwidth]{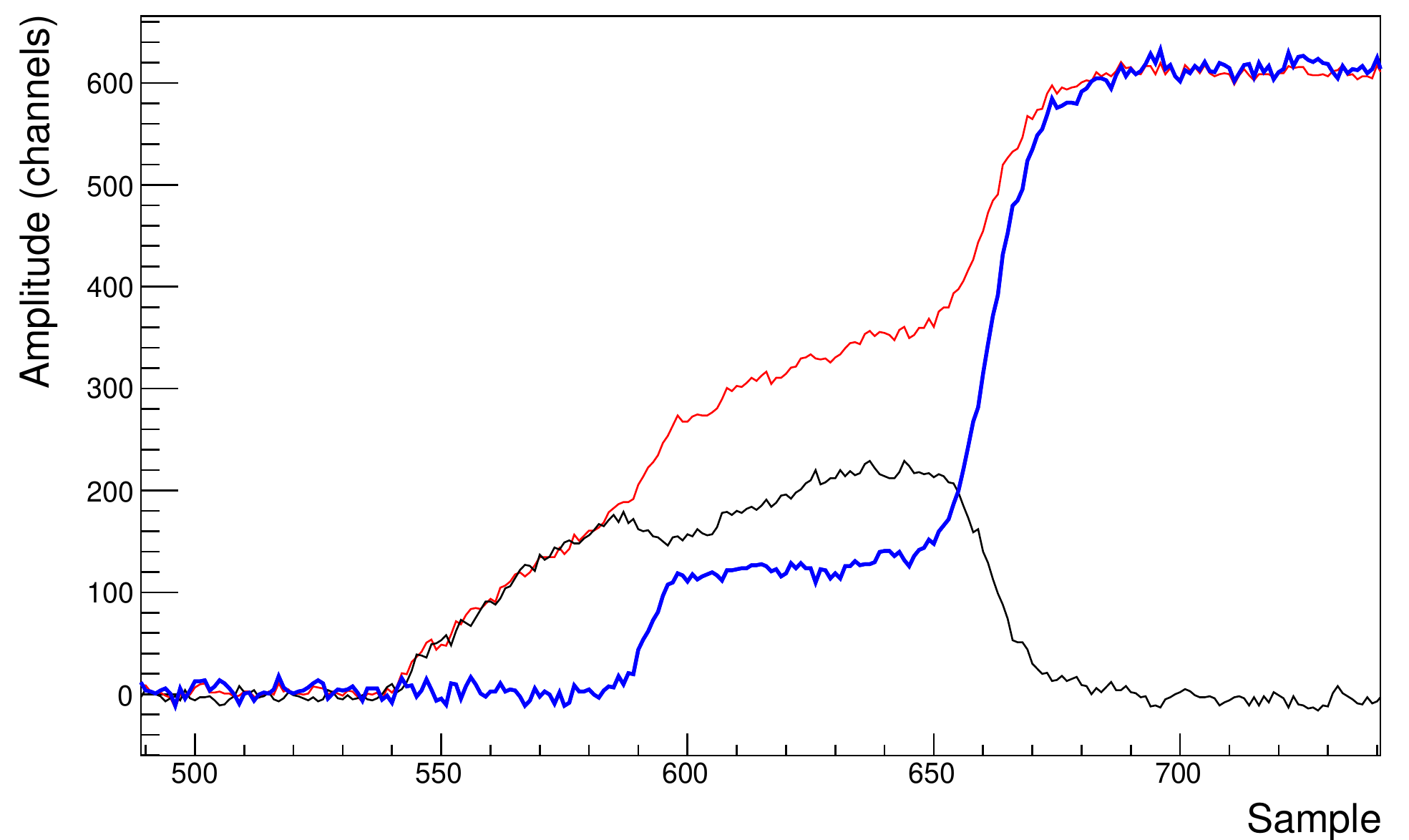}
 \caption{Pulses from a multi-site gamma event. Red: CA pulse. Black: NCA pulse. Blue: Difference pulse. The event shown is a 1.8\,MeV interaction with an ERT value of 78 samples.
}
 \label{fig:MSEevent}
\end{SCfigure}

A cut to remove LSEs will remove some uniformly-distributed gamma events simply because a fraction of those events
occur close to a lateral surface. However, a cut based on ERT can also be expected to remove many multi-site events (MSEs).
Many gamma events involve multiple effectively-simultaneous interactions within the detector.  
These MSEs have their own pulse distortion signature, with multiple steps in the difference signal
(see figure ~\ref{fig:MSEevent}) which can also produce high ERT values. 
Much of the high-ERT tail in the gamma distribution seen in figure~\ref{fig:cutDiscrim} can be attributed to MSEs.
Indeed this tail rises above the alpha distribution at the highest ERT values. 
The energy range depicted, 1.7 to 2.0\,MeV, avoids peaks in the gamma spectrum, so 
most of these events are Compton scatters, either single- or multi-site. 

Since the signal events sought by the COBRA experiment are beta interactions, which rarely produce MSEs, the accidental identification of MSEs as LSEs is not considered a disadvantage for discriminating background. 
It does however complicate measuring signal efficiency for an LSE cut, since it will differ for gamma and beta events if MSEs are not removed from consideration.


\section{LSE identification in the LNGS array}

\begin{figure}[h]
 \includegraphics[width=1.0\textwidth]{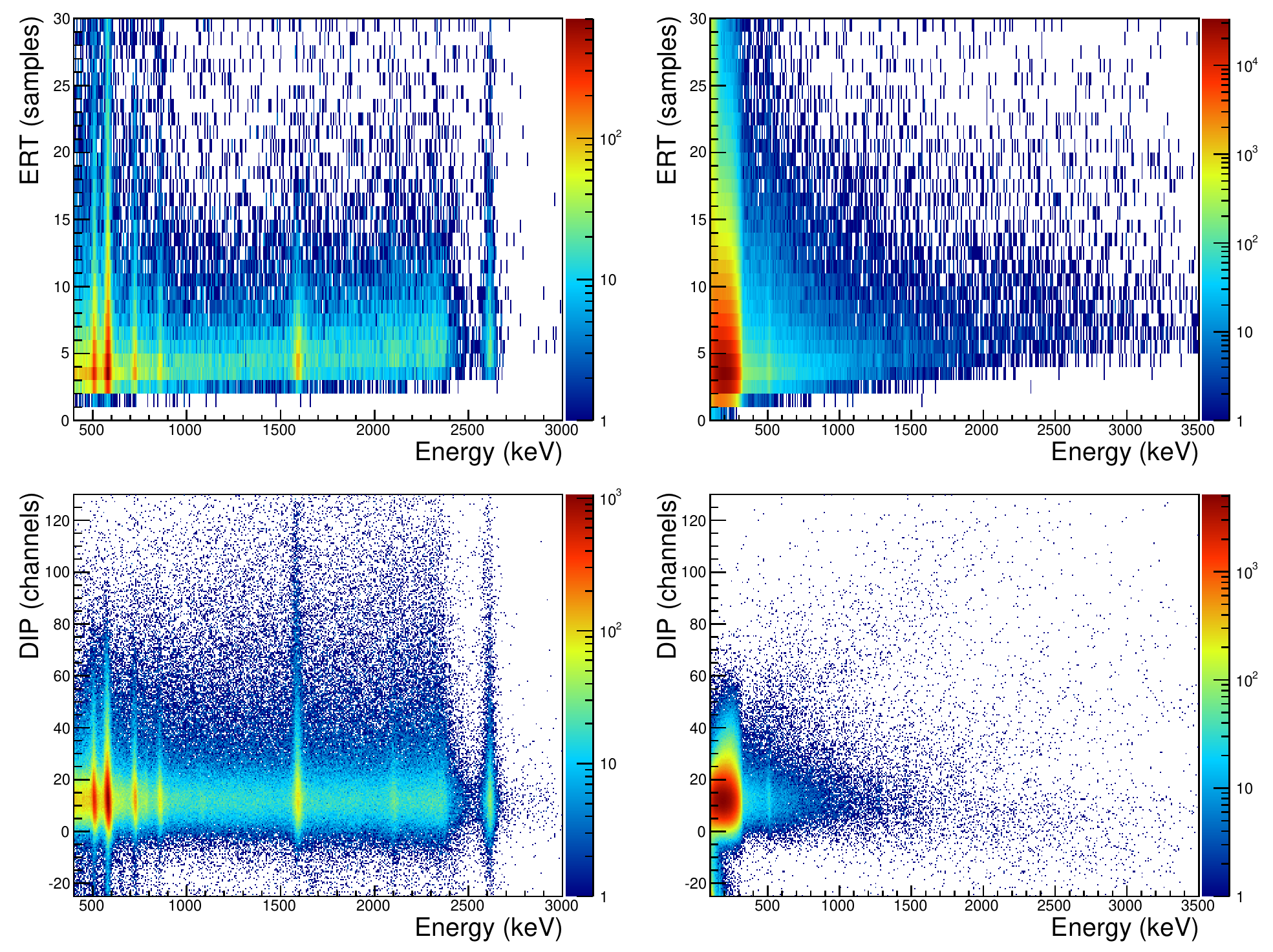}
 \caption{LSE identification quantities
 versus energy for data from the LNGS detector array. Left: from a \isotope[228]Th calibration. Right: from low-background physics runs. The region of interest for the neutrinoless double-beta decay of \isotope[116]Cd is near 2800 keV. 
 In all of these figures events have been pre-selected to remove events near the cathode and anode surfaces, based on the 
 calculated interaction depth along the longitudinal axis.
} 
 \label{fig:LNGSertdipVSe}
\end{figure}

The LSE identification quantities described in this paper have been used 
to develop analysis cuts for the array of detectors deployed
in the low background setup at LNGS for the COBRA experiment. 
Figure~\ref{fig:LNGSertdipVSe} shows the distribution of the ERT and 
DIP quantities versus energy for data from the LNGS detector array.
Each plot combines data from 30~detectors. 
Since surface events near the anode and cathode surfaces are easily 
identified based on the calculated interaction depth along the 
longitudinal axis, they have been removed from these plots. Thus only 
central events and LSEs are represented. The left plots in the 
figure show data from a calibration run using a \isotope[228]Th gamma source. 
These plots are qualitatively similar to the left plots in figure~\ref{fig:ertdipVSe}.

The right plots in figure~\ref{fig:LNGSertdipVSe} show data from the detector array running in the low-background mode used to search for rare decays. 
Most events in these plots are at energies below 330\,keV, arising from the intrinsic beta decay background of $^{113}$Cd. Three gamma peaks have
been identified in the background which are slightly visible in these plots, at 511\,keV and 1275\,keV apparently coming from \isotope[22]Na, and 1461\,keV from 
\isotope[40]K. Both the $^{113}$Cd background
and the gamma peaks have LSE-quantity distributions similar to those seen in the \isotope[228]Th data. To date (after the accumulation of 82.3\,kg$\cdot$days of exposure) 
no significant population of gamma events has been found in the background at energies above 1500\,keV. 
Indeed the ERT and DIP values of the high-energy background do not fall into the chararacteristic band seen for gamma and beta events, but rather
take on wide distributions. This suggests that the high-energy background is largely comprised of LSEs.

The LSEs seen in the high-energy background are believed to arise mostly from contamination of the detector surfaces
and surrounding hardware by alpha-emitting isotopes. Such contamination has been seen on the cathode surfaces, 
where alpha peaks are evident. In particular, a peak at 5.2 MeV is 
identified as coming from $^{210}$Po, which is a daughter of $^{210}$Pb. $^{210}$Pb is the only long-lived daughter in the radon decay chains, 
and can be expected to contaminate the surfaces of detectors and other equipment during testing and storage in environments where radon is present. 
Thus it is a reasonable assumption that all surfaces of the detectors have some $^{210}$Pb contamination. Unlike the cathode
surface background, no peaks are seen in the high-energy LSE background, but this is expected from the detector construction. While at the cathode 
the crystal is only covered by approximately 100\,nm of metal, on the lateral surfaces there is a coat of encapsulating lacquer, some 40\,$\mu$m thick
on average (though far from uniform). Thus the energies of alpha particles emitted at the surface of the paint will be strongly attenuated before 
they enter the crystal to produce LSEs. 

\begin{figure}[h]
 \includegraphics[width=1.0\textwidth]{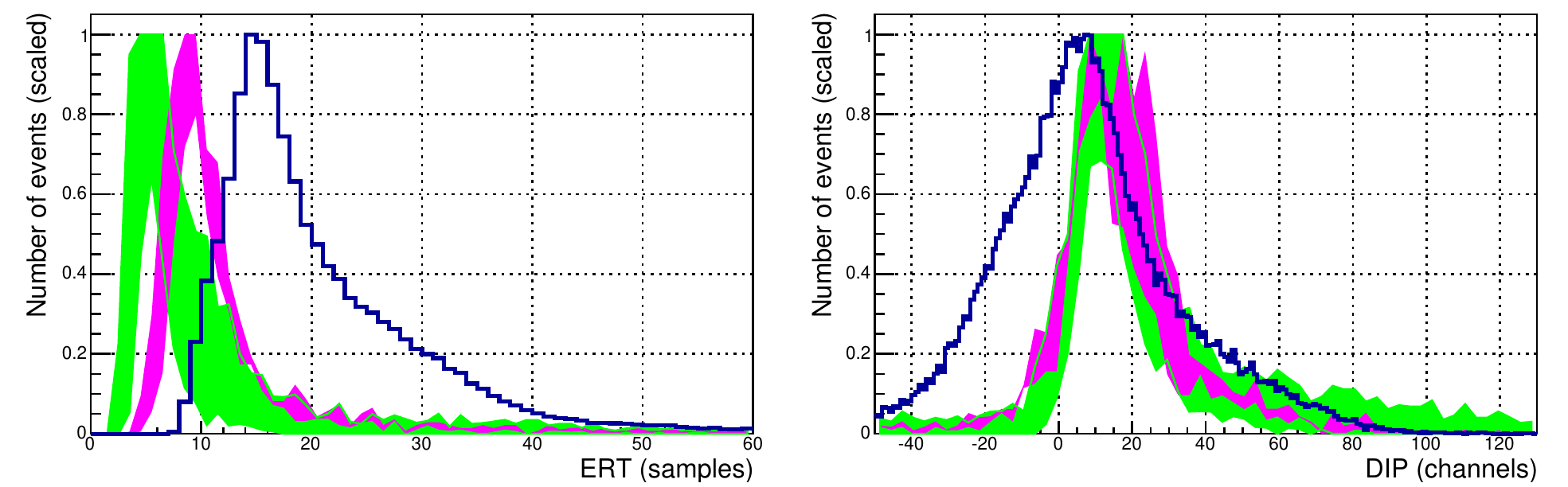}
 \caption{Distribution of LSE identification quantities for gammas in an energy range
of 1.7-2.0\,MeV, comparing the 30 high-resolution LNGS array detectors (green), the 2 lower-resolution LNGS
array detectors (magenta), and the test detector (blue). The thickness of the green and magenta 
curves indicates the full range of values for the detectors included.   
Left: ERT distributions. 
Right: DIP distributions. } 
 \label{fig:DO-LNGScompare}
\end{figure}

Significant differences exist in the gamma distributions of the LSE identification quantities between the test detector described in the previous sections 
and the LNGS array detectors described in this section.
While the test detector is of the same design as those
in the LNGS array, it is of a different batch and is a lower quality crystal. 
The LNGS array consists of 32 detectors, two of which are from an older, lower-resolution batch.
Figure~\ref{fig:DO-LNGScompare} shows the 1.7--2.0\,MeV gamma distributions of the LSE identification quantities comparing the test detector
to the two batches of detectors from the LNGS array. The two most notable features of the test detector that differ
from the array detectors 
are a higher characteristic value for ERT and a prominent tail in the DIP distribution extending to negative values. 
Both of these differences can be attributed to slower timing characteristics in the difference pulses from the test detector.    

Variations in the ERT and DIP characteristics among the detectors in the array are comparatively small.
To develop analysis cuts to identify and veto LSEs in the LNGS detector array, a common ERT threshold has been set for the 30 detectors in the 
higher-quality batch, with a different threshold chosen for the two detectors in the lower-quality batch. A common DIP threshold is set for all 32 detectors.  
The thresholds are chosen
so that approximately 90\% of uniformly distributed single-site events are retained for each cut, for a total signal efficiency of approximately 80\%. 
With cathode- and anode-surface events removed, 1978 events are present between 2\,MeV and 4\,MeV in the current 82.3\,kg$\cdot$days of low-background data. 
The combined LSE cut removes
75.0\% of these events. 53.1\% of the events have above-threshold ERT values, and 23.0\% have above-threshold 
DIP values -- thus these events are
categorized as CA-side and NCA-side LSEs, respectively. 
1.1\% of the events have above-threshold values for both ERT \emph{and} DIP values. These appear to be NCA-side events with unusually long rise times, perhaps due to reduced electric fields along certain charge paths.

It is possible that many of the high-energy background events that have not been removed by this LSE cut are nevertheless LSEs that have 
simply not been identified as such. Along the lateral surfaces of the detector there are two edges where the CA- and NCA-sides meet (see 
figure~\ref{fig:cpg}). For LSEs near these edges there should be some transition between pulses with high ERT values and pulses with high
DIP values. Indeed the two types of distortions are likely to cancel one another, so that some LSEs in these regions escape identification. 
Preliminary detector simulations indicate that the region of effect lies mostly on the NCA sides, which helps explain why fewer LSEs are identified 
as NCA-side than CA-side by the LSE cut. 
Laboratory exploration of the behavior of the quantities for LSEs at different positions along the lateral perimeter, as well as different 
positions along the longitudinal axis, are ongoing, with the intent to fully characterize the pulse shape properties of LSEs at all positions
on the lateral surfaces. 

In addition to calculating the background reduction of the LSE cut, for a rare event search analysis it is also necessary to precisely calculate
the signal efficiency of the cut.
 For the LNGS array this has been done using \isotope[228]Th calibration data.  Based on
present calculations the
efficiency has been found to decrease slowly with energy, by about 1.7\% per 1000\,keV. This can be expected from the behavior of DIP. 
Among central events some will have larger intrinsic (i.e.\ not noise-based) DIP values than others, and since DIP is roughly proportional to energy, more
high-energy central events will be accidentally identified as LSEs than low-energy central events. More significant is the variation of the
efficiency across different detectors, arising from the differences in the ERT and DIP characteristics. These variations can arise from differences in crystal quality, precision of electrode patterning, applied bias
voltages, and noise conditions. At the primary region of interest near 2.8 MeV, the mean signal efficiency of the LSE cut has been calculated at 81\%, spanning a range of 70\% to 90\% across detectors.   
The calculation of the efficiency is complicated by the presence of MSEs in gamma event populations, as explained previously. Development of efficient MSE identification for COBRA detectors is in progress,
intended primarily to reduce potential MSE background but with a secondary benefit of improving the calculation of the LSE cut efficiency. 

Greater understanding of the LSE discrimination technique, from further laboratory tests and better MSE identification, should improve the performance of LSE cuts developed for future use
in the COBRA experiment.  The cut can be tuned on a detector-by-detector basis, and perhaps defined in an energy dependent way, to balance signal efficiency and background reduction
in order to optimize the sensitivity to a neutrinoless double beta decay signal.
Smoothing techniques such as averaging or interpolation could be added to the algorithms for calculating ERT and DIP, which may improve their discrimination power.

\section{Conclusion and outlook}
Deformations in the weighting potentials near the lateral surfaces of CPG detectors result in deformations in the pulse shapes produced by events occurring
at the surfaces, allowing for their identification. 
An algorithm developed to measure those distortions has been successfully tested with laboratory measurements using alpha radiation as a source of surface events. 
An implementation of the LSE discrimination technique applied to the LNGS low-background physics data 
of the COBRA experiment reduces the background by 75\% in the energy region of interest, with an estimated loss of 19\% of the uniformly-distributed signal, 
increasing the signal to background ratio by a factor of 3.2. 
Open questions regarding the edges of the lateral surfaces, where an NCA side meets a CA side, are currently being investigated, 
as well as an interaction depth dependence of the lateral surface event identification quantities. 

The quantitative details of LSE identification vary noticeably across different detectors used by the COBRA experiment, largely due 
to differences in crystal quality. In addition the results presented here are specific to the CPG design and operation mode currently in use by the COBRA 
experiment. For detectors with a different grid design, or with a guard ring held at a fixed electric potential, the LSE discrimination technique
may require significant modification.

\section*{Acknowledgments}
The COBRA Experiment would like to thank the Istituto Nazionale di Fisica Nucleare for the use of their laboratory and infrastructure at the Laboratori Nazionali del Gran Sasso, as well as the Deutsche Forschungsgemeinschaft for their support.

\section*{References}

\bibliography{cobra_lateral_surface_event_discrimination}
\bibliographystyle{ieeetr}

\end{document}